\documentclass{svproc}
\usepackage{url}
\usepackage{graphicx}
\usepackage{amsmath,amssymb,amsfonts}

\begin{document}
\mainmatter              
\title{An Agent-based Cloud Service Negotiation in Hybrid Cloud Computing}
\titlerunning{An Agent-based Cloud Service Negotiation}  
%
\author{Saurabh Deochake\inst{1} \and Debajyoti Mukhopadhyay\inst{2}}
\authorrunning{Deochake and Mukhopadhyay} 
%
\tocauthor{Saurabh Deochake, Debajyoti Mukhopadhyay}
\institute{Twitter, Inc., Boulder CO 80302, USA,\\
\email{sdeochake@twitter.com}
\and
WIDiCoReL Research Lab, Mumbai University, Mumbai 400032, India\\
\email{debajyoti.mukhopadhyay@gmail.com}}

\maketitle              

\begin{abstract}
With the advent of evolution of cloud computing, large organizations have been scaling the on-premise IT infrastructure to the cloud. Although this being a popular practice, it lacks comprehensive efforts to study the aspects of automated negotiation of resources among cloud customers and providers. This paper proposes a full-fledged framework for the multi-party, multi-issue negotiation system for cloud resources. It introduces a robust cloud marketplace system to buy and sell cloud resources. The Belief-Desire-Intention (BDI) model-based cloud customer and provider agents concurrently negotiate on multiple issues, pursuing a hybrid tactic of time and resource-based dynamic deadline algorithms to generate offers and counter-offers. The cloud marketplace-based system is further augmented with the assignment of behavior norm score and reputation index to the agents to establish trust among them.
\keywords{Automated Negotiation, Intelligent Systems, Cloud Computing, Multi-agent Systems}
\end{abstract}
\section{Introduction}
Negotiation is a mechanism in which different agents concede to an agreement based on a joint future venture \cite{p1}. For a technology organization, expanding the on-premise infrastructure is no longer limited to expanding the hardware footprint in the data centers. The organization may benefit from expanding their infrastructure into a multi-tenant system like public cloud, embracing a hybrid cloud infrastructure model. The recent studies of e-negotiation can be applied to such unique scenarios of hybrid cloud computing. A well-known example is Amazon AWS EC2 Marketplace that allows customers to bid for EC2 Spot Instances virtual machines. To implement a robust cloud marketplace system, the transactions involving exchange of offers and counter-offers conceive the alliances that could be either ad-hoc or permanent. This agenda necessitates a strict coherence to an appropriate set of protocols and a threshold for every negotiation agreement between a consumer and a provider agents\cite{p2}. An agent is an encapsulated computer system based on an unknown external environment with the ability of elasticity and self-governing actions in that environment in order to meet its designed objectives. The Belief-Desire-Intention (BDI) \cite{book1}, a software model designed to program intelligent agents, is perhaps that most studied model for bounded rational systems. BDI model closely associates with practical human thinking and abstract logical reasoning enabling rational BDI agents perform complex tasks in a dynamic environment using real-time reasoning frameworks like Procedural Reasoning Systems (PRS). \par
The negotiation agenda possesses a multitude of issues that need to be fulfilled in order to reach an agreement. In the scenario of hybrid cloud computing negotiation, the agenda may consist of a single object such as price of a virtual machine resource or the issues may be multiple such as offered memory size, disk storage size and the operating system image for a virtual machine. The transactions become even more multifaceted when they involve issues related to non-functional attributes like cloud service-level agreement (SLA). A cloud customer may concurrently negotiate with multiple cloud providers on the issues for a product that are the most important to the customer. However, before commencing a negotiation round, all stakeholders must agree upon the issues in the negotiation to form an alliance. Alliances are formed when multiple parties join hands together in pursuit of a command business goal.\par
The paper presents a multi-party multi-issue negotiation system for BDI-based agents that adapt to advancements from each round of negotiation and make appropriate counter-offers in a cloud marketplace to reach an agreement. The paper is structured as follows: section 2 discusses previous work, section 3 showcases the model, section 4 presents the algorithmic approach, section 5 presents future work, and finally, we conclude this paper in section 6.

\section{Related Work}
BDI-based programming paradigm \cite{book1} is permeated with philosophical attitudes viz. Beliefs, Desires and Intentions. Beliefs represent the agent's current knowledge of the outside world. Desires represent the objectives that the agent may carry out based on the current Beliefs. Finally, Intentions are the ruminative state of the agent. A few shortcomings of original BDI-based agents such as lack of learning and adaption, and multi-agent competence as showcased in \cite{p4} were attempted to alleviate, for example, in \cite{p5} that showcases distributed Multi-Agent Reasoning System (dMARS) architecture that couples PRS with BDI architecture to enable distributed BDI agents to learn and adapt. Additionally, the work in \cite{p6} extends the original BDI agent architecture to facilitate multi-agent paradigm with shared beliefs using distributed shared memory and network messages. With the work that has been done in the area of learning and multi-agent system capability, BDI agents can be used to solve complex problems with limited information operating in a dynamic environment \cite{p7}. Our paper relies on BDI agents' competence to adapt to constantly changing environment when negotiating complex cloud service issues.\par
With the advent of evolution in cloud computing \cite{p8}, an organization looking to expand their IT infrastructure to cloud has multiple choices for cloud service providers. In order to choose the best deal, it must negotiate with multiple cloud providers to select the best possible outcome. To the authors' best information, most of the study has been focused on negotiating the SLAs in cloud computing \cite{p9,p10,p11}. The paper in \cite{p9} discusses negotiating cloud service SLA for grid computing where the process owner specifies service-level requirements from business process perspectives and selects a cloud provider. While this approach works well negotiating non-functional parameters like SLAs from a pre-elected opponent, it does not discuss negotiating functional parameters like cloud service's specifications with multiple parties in a cloud marketplace scenario. To the authors' best belief, cloud agency-based approach such as mOSAIC \cite{p10} adopts a software-based multi-cloud negotiation system. It proposes a middleware-like cloud agency model that negotiates on behalf of a cloud customer. However, this approach does not consider a fine-grained control such as provisioning a concurrent multi-party multi-issue negotiation. Our approach proposes a marketplace-based negotiation system where each BDI-based agent sets a fine-grained control on the issues that it values the most in form of weights. \par Traditional negotiation exhibits pre-defined strategy that works well with an opponent. For instance, in \cite{p2,p12}, the agent selects a predefined negotiation strategy that works well with a specific negotiation scenario using the game theory principles seeking a win-win outcome. However, this approach forces both buyers and sellers to lose their utility in reaching the agreement. In our work, BDI agents adapt to each negotiation round, selecting appropriate tactic to reach an agreement within a fixed time period. While \cite{p13} discusses a bilateral multi-issue negotiation system where a buyer and seller negotiate on multiple issues concurrently, buyer and seller must reach an agreement on every issue for a consensus. This leads to over-negotiating between a buyer and seller that potentially continues for extended period of time. While Belief-Goal-Plan (BGP) based multi-strategy approach \cite{p14} makes it easier for an agent to select the best-working negotiation strategy to counter the opponent, it does not take into account multi-lateral concurrent negotiation on multiple issues. Furthermore, negotiation may not terminate within a fixed time period if both buyer and seller have a set of mutually exclusive issues. Our approach is inspired by Faratin's time and behavior dependent negotiation strategy, selecting a set of issues with least associated weight to concede to the opponent. A similar attempt has been performed in \cite{p15}. Allowing negotiation between two parties on mutually exclusive issues would inundate the negotiation system with quotes with few interested opponent agents. In our approach, our framework allows each party to select the issues they deem the most important and the opponent is matched for negotiation based on the shared mutual interests to reduce the overhead on the negotiation engine.

\section{Proposed System Architecture}
As showcased in Fig. 1, our proposed system architecture is composed of a cloud marketplace and cloud agents. The cloud marketplace facilitates negotiation among multiple agents on a certain subset of issues. It also provides the marketplace for buyer and sellers of cloud services to advertise their products and procure the services.
\begin{figure*}
  \includegraphics[width=\textwidth]{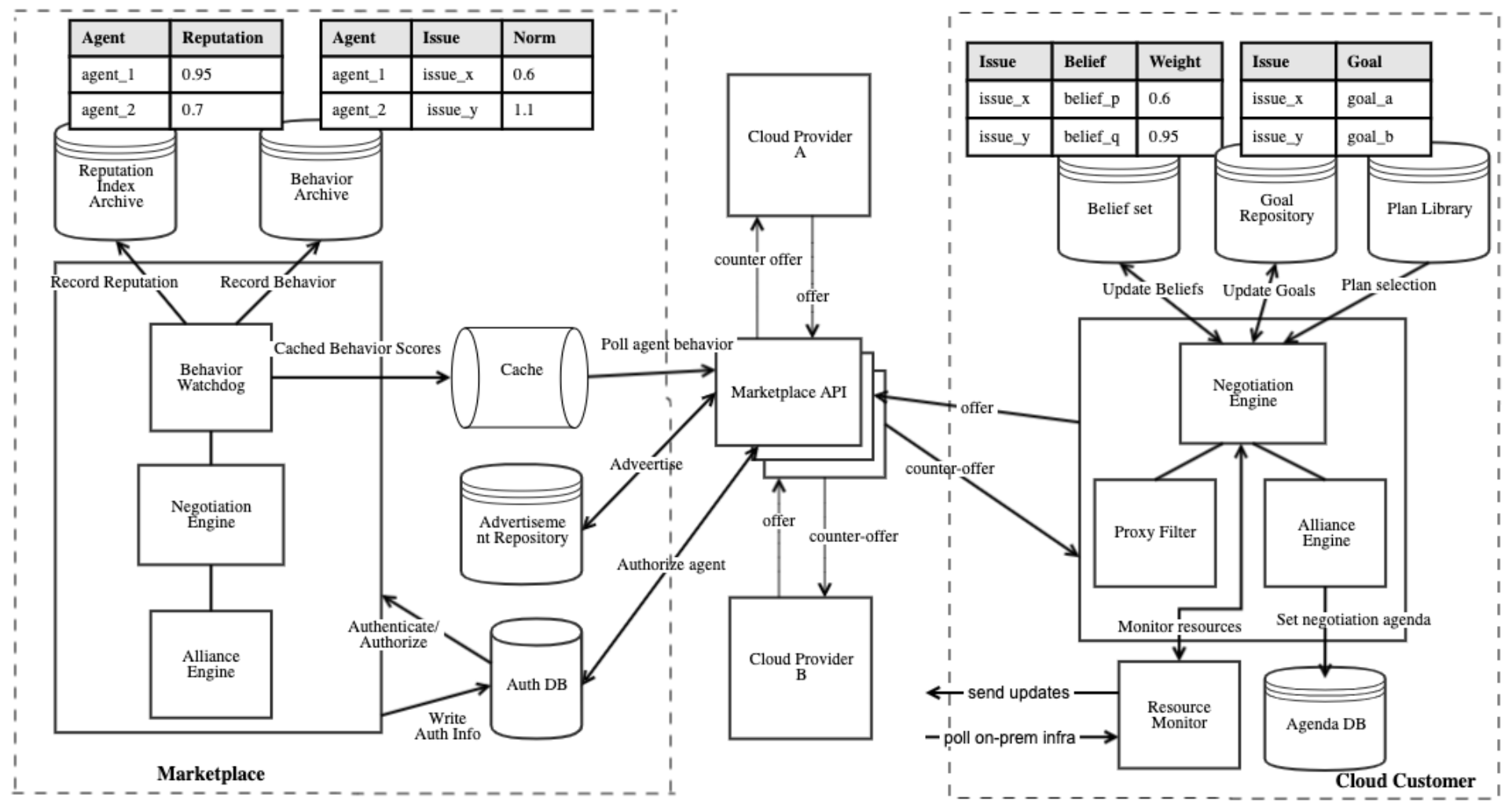}
  \caption{Proposed System Architecture - An Agent's View}
\end{figure*}
\subsection{Cloud Marketplace}
The Cloud Marketplace is a central distributed system where cloud providers and customers submit their advertisements. The negotiation system and the associated web services are implemented using WSDL (Web Service Description Language), XML and SOAP (Simple Object Access Protocol). The Cloud Marketplace back-end components can be interacted via a RESTful Marketplace API service after a successful authentication.\par
Advertisement Repository stores the agent and product information sent by cloud providers and customers. It assists in agent and product discovery by querying the parameters like agent ID, product ID and corresponding issues. The cloud customer agent that is looking to procure a service from a cloud provider will send a Request for Quote (RFQ) with its information, product information and a certain criteria such as Reputation Index for an agent. Other services poll Advertisement Repository for match-making and statistical purposes. \par
Cloud provider and customer may have their own agenda on how important they deem individual issues of a product. Expensive compute resources in distributed multi-threaded transactions \cite{p16} necessitates the provisioning of a feature that would limit the futile concurrent negotiation on issues that are mutually exclusive for both agents. Alliance Engine is essentially a match-maker between a cloud provider agent and a customer agent that consistently polls the Advertisement Repository database for issues of mutual interest and other parameters in RFQ like Reputation Index for a match. This service then forwards a successful match to the involved agents via Negotiation Engine with a message to commence the negotiation. \par
Negotiation Engine is a pivotal service where all transactions between the two agents occur. After an alliance is formed, a \texttt{commenceNegotiation} message, along with a timestamp in UTC, is sent to both the agents. The cloud provider agent starts the negotiation by sending an offer to a customer agent. The rounds of offers and counter-offers continue between the agents for a threshold of time. Negotiation system also keeps track of all the transactions between agents and gathers statistics about the negotiation process and agents. These statistics are then fed into Behavior Watchdog service that assigns behavior and reputation indices for the agents.\par
A cloud customer agent may prefer negotiating with a cloud provider that provides better SLAs, cost effective resources and faster agreements with its customers. A cloud provider may have similar expectations. Based on the statistical data gathered by Negotiation Engine, Behavior Watchdog service calculates Reputation Index, a rational number value- \textit{R}, where 0 $\leq$ \textit{R} $\leq$ 1, stored in a database named Reputation Index Archive. More trustworthy and reputable agents possess higher value, and tend to generate more agreements than lesser reputable agents. Similarly, an agent could use the information about common behavior of its opponent to draft a negotiation strategy. Based on the negotiation messages, Behavior Watchdog also curates a generalized Behavior Norm index, \textit{B} for an agent that indicates how an agent generally behaves in negotiations. Behavior Norm index is also a rational number value that decides the nature of the agent as mentioned in (1) for the strategy \textit{S}. 
\begin{equation}
    \mathcal{S} = 
    \begin{cases}
    B_a < 1, \text{where agent a tends to be headstrong}\\
    B_a = 1, \text{where agent a tends to be linear}\\
    B_a > 1, \text{where agent a tends to be conceder}\\
    \end{cases}
\end{equation}

\subsection{Cloud Agent}
Each BDI-based cloud agent has three relational database systems that store agent's beliefs, goals and plans. An agent's beliefs, stored in Beliefset, are the knowledge about the current negotiation round on an issue \textit{i} and are constantly updated after each round. The agent assigns weights to every issue in the negotiation, indicating the importance of that issue, such that $\sum_{i=1}^{n} W_i = 1$. A goal can be an activity of trying to reach an agreement for an utility of an issue based on the belief gained from the previous round of negotiation. Like beliefs, goals are updated every round of negotiation. In terms of negotiation, a plan would indicate that the agent has started to execute an offer, a counter-offer or termination of the negotiation. Frequently occurring scenarios of plan creation are stored in a database named Plan Library.\par
A negotiation between two agents requires a strict adherence to the agreed issues and a time threshold. Receiving a counter-offer from an opponent after the time threshold is passed would be an illegal transaction. Additionally, to secure an agent from a malicious attempt to disrupt its negotiation with other agents, it becomes crucial to accept only those transactions that are valid. Proxy Filter filters illegal messages pertaining to a negotiation if a transaction a counter-offer is outside of current negotiation space.\par
An agent may enter negotiation to buy or sell its product with multiple  agents at once reaching an agreement with an opponent that offers a higher utility. It becomes essential for an agent to be aware of every negotiation space that it is participating. Alliance Engine service is responsible for curating the information about a negotiation space with opponent agents, storing the metadata of all active negotiations in Agenda DB.\par
Negotiation Engine service is the heart of the agent software. This service performs essential transactions such as generating offers and counter-offers, evaluating those offers and selecting appropriate negotiation strategy. It is also responsible for submitting advertisements and RFQs for the issues to the Cloud Marketplace API. After each round of negotiation, Negotiation Engine service also updates Beliefset, Goal Repository and Plan Library for the agent.\par
Resource Monitor constantly polls the on-premise resource availability by means of a monitoring service. Based on the dynamic availability of the resources, the agent will ensure that it selects an appropriate negotiation strategy. If Resource Monitor alerts the agent that the resource it is looking to purchase is running short, the agent will choose to be aggressive in reaching an agreement as soon as possible.

\section{Negotiation Process}
Before the start of the actual negotiation process, a cloud agent submits its desires such as weight of an issue (W\textsubscript{i}), a range of acceptable cost value (C\textsubscript{i}) to each issue, i as mentioned in (2) and a threshold value for the negotiation time period (t\textsubscript{max}) such that t\textsubscript{min\textsubscript{i}} $\leq$ t\textsubscript{i} $\leq$ t\textsubscript{max\textsubscript{i}}. The utility value for each issue submitted by the agent is then calculated as per (2).
\begin{equation}
  U = \sum_{i=1}^{n} C_i W_i \text{  where } \sum_{i=1}^{n} W_i = 1 
\end{equation}
When the negotiation rounds start, there are multiple ways of how an agent records its response, \textit{R}. For a transaction between agents a and b, an agent may acquire the offer for an issue i (O\textsubscript{i}\textsuperscript{t\textsubscript{n-1}}) if the measured utility of that offer is greater than or equal to the utility of the counter-offer. Otherwise, the agent sends a counter-offer with a modified cost value for an issue i. As per the protocol, if an counter-offer is received beyond the threshold value of the negotiation time period, the agent would terminate the negotiation round as per (3).

\begin{equation}
    \mathcal{R} = 
    \begin{cases}
        acquire(a, b, O_{i}^{t_{n-1}}), & U_i({O_i}_{a \rightarrow b}^{t_{n}}) \leq U_i({O_i}_{b \rightarrow a}^{t_{n-1}})\\
        terminate(a, b), & t_\textit{co} > t_\textit{max}\\
        counter(a, b, {O_i}_{a \rightarrow b}^{t_{n}}), & \text{otherwise}
    \end{cases}
\end{equation}

Once an agent enters negotiation process, the goal of the agent is maximizing the utility for all the issues in the product that the agent deems important. In order to do so, the agent should emphasize on counter-offer generating strategies. As showcased in \cite{p17,p18}, the agent may use dynamic time-dependent tactic where offers vary as the deadline approaches, based on a parameterized time function. On the other hand, dynamic resource-dependent tactic enables agents vary their offers based on the quantity of resources. In our model, cloud customer and cloud provider agents use a hybrid time and resource-dependent deadline-based tactics to generate counter-offers. As mentioned in \cite{p17}, time-dependent tactic is governed by a time-based function \textit{f(t)} where an agent must conclude the negotiation within the threshold value of t\textsubscript{max} such that 0 $<$ t $\leq$ t\textsubscript{max}. A counter-offer \textit{O} for issue i from cloud agent a to b at time t\textsubscript{n} is measured as per (4).

\begin{equation}
    {O_i}_{a \rightarrow b}^{t_{n}} = \\
    \begin{cases}
    {min_i}_a + {{f_i}_a}(t_n)({max_i}_a - {min_i}_a) \text{, if cost decreases}\\
    {min_i}_a + ({1-{f_i}_a}(t_n))({max_i}_a - {min_i}_a) \text{, if cost increases}\\
    \end{cases}
\end{equation}

Furthermore as showcased in \cite{p17}, the time-based function that depicts the agent behavior is based on fulfilment of following constraints, \textit{L} as per (5).

\begin{equation}
    L = \\
    \begin{cases}
    0 \leq {{f_i}_a}(t_n) \leq 1 \text{, offer range}\\
    {{f_i}_a}(0) = {{C_i}_a}(0) \text{, initial cost belief at time 0}\\
    {{f_i}_a}(t_{max}) = 1 \text{, at the time threshold}\\
    \end{cases}
\end{equation}
After a round of negotiation, if a counter-offer sent by opponent agent does not fall within the acceptable utility range. A new offer is generated based on a concession rate, $\lambda$ for the cost value of the issue as shown in (6).

\begin{equation}
    \lambda = \frac{{O_i}_{b \rightarrow a}^{t_{n}} - {O_i}_{b \rightarrow a}^{t_{n-1}}} {{O_i}_{b \rightarrow a}^{t_{n-1}} - {O_i}_{b \rightarrow a}^{t_{n-2}}}
\end{equation}

In the hybrid time-resource dependent counter-offer tactic, a hardheaded tactic is applied near the negotiation threshold or if opponent's resources run short. That is, concessions are made near the end of the time threshold, t\textsubscript{max} or resource threshold, r\textsubscript{max}, whichever is smaller. On the other hand, a conceder tactic is applied if an agent is desperate to reach a deal quickly by applying concessions to its offers starting early in the negotiation rounds. An agent may start conceding in the early rounds of the negotiation and then depending on the resources, it may choose to alter the tactic to become more hardheaded. If the agent finds that the opponent is practising a linear conceding tactic then the agent will try to match it, changing the counter-offers only slightly.\par To calculate the concession rate, $\lambda$, an agent must wait until the third round of negotiation, if it finds a better utility offer before third round, the agent will accept that offer. Beyond third round, negotiation engine service calculates the concession rate by evaluating past offers as per (6). Concession rate of $\lambda$ $>$ 1 signifies opponent's conceding strategy and the agent will imitate opponent tactic. On the other hand, concession rate of $\lambda$ $<$ 1 conveys opponents headstrong tactic, the agent chooses conceding tactic to facilitate faster agreement. Agent tactic is unchanged for the linear concession rate of $\lambda$ $=$ 1.

\section{Future Work}
The paper discussed a negotiation model ensuring concurrent negotiation transactions with multiple opponent agents. The paper focuses on a single owner of IT infrastructure looking to scale their infrastructure into the cloud. However, the authors believe that this is not indicative of all organizations and their infrastructure setup. We further intend to examine more complex use cases allowing individual microservices in the infrastructure to automatically negotiate with public cloud providers and scale themselves. We plan to study the possibility of assigning an intelligent agent to an individual service and explore how those multiple agents together form multi-agent system and scale the overall infrastructure. Another aspect of future work refers to extending this negotiation system for multi-agent behavior \cite{p19} and communication \cite{p20} using paradigms like distributed shared memory and Law-governed Interaction, respectively.

\section{Conclusion}
This paper presented an enhanced cloud marketplace-based automated negotiation system for hybrid cloud computing that exhibits multi-party, multi-issue negotiations. To reduce the compute overhead incurred in parallel negotiation, it introduces forming the alliances with converging agenda among BDI agents. The cloud marketplace system ensures that an agent may choose to transact with a certain set of reputable agents. The model also enables agents to craft their negotiation tactic using hybrid time-resource deadline algorithm. The paper paves way toward an automated scaling of an on-premise infrastructure into public cloud. While the current approach presented in the paper focuses on a single agent, further study is underway to augment the current work with the provisions of distributed learning and interaction in a multi-agent system.

%
%

\end{document}